# Global Platform for Rich Media Conferencing and Collaboration

David Adamczyk, David Collados, Gregory Denis, Joao Fernandes, Philippe Galvez,
Iosif Legrand, Harvey Newman, Kun Wei
*California Institute of Technology*
*1200 East California Boulevard, Pasadena California 91125*

The next generation of HENP (High Energy and Nuclear Physics) experiments largely involve dispersed collaborative environments, supporting point to point and multipoint videoconferencing and application sharing. Since 1995, The "Virtual Rooms Videoconferencing System" (VRVS) is being developed by Caltech (California Institute of Technology) in order to provide a low cost, bandwidth-efficient, extensible means for videoconferencing and remote collaboration over networks within the High Energy and Nuclear Physics community and with extensions also to other research communities. VRVS supports full connectivity to Access Grid through the VRVS AG Gateway, where the VRVS users can easily connect to the Virtual Venues or any multicast videoconference. In February, the new VRVS version 3 has become the production system. Its key points are robustness and scalability: thousands of users connected to hundreds of meetings at the same time. The new design of the system enables a comfortable integration of new tools, clients, emerging standard, and a backup routing path between Europe and America. VRVS continues to expand and implement new digital video technologies. Further aspects include among others: SIP, Pocket PC platform support and services (based on the MonALISA[1] project) to monitor real-time activity, oversee and manage the whole distributed system in a dynamic way.

## 1. INTRODUCTION

The Caltech "Virtual Rooms Videoconferencing System" (VRVS) (http://www.vrvs.org) has become a standard part of the toolset used daily by a large sector of HEP, and it is used increasingly for other DoE-supported programs. It has also attracted substantial interest in diverse fields of science and engineering outside the HEP. The relationship with Internet2 and other international organizations has been consolidated.

The development of the next generation of the VRVS system is currently underway; details are described in latter sections of this document. As a software-based collaborative infrastructure, VRVS is now recognized as a great application capable of scaling to provide future collaboration services throughout these organizations.

## 2. THE VRVS PROJECT

The VRVS system is managed by the Caltech CMS group. Since last year, the number of multipoint collaborative sessions (national and international) using VRVS increased by 180% mainly pushed by the successful transition to the new version, where a nearly unlimited set of virtual rooms is available. This large adoption rate confirmed the need within the Research and Academic communities for easy-to-use collaborative tools with high performance.

As part of the Internet2 Commons initiative, Internet2 and the VRVS Caltech Team continue to deploy a series of VRVS servers (currently 11), known as reflectors, over the Abilene Backbone and Internet2 Universities. The objective is both to provide better performance for existing VRVS users, and to facilitate access by new users. VRVS uses the Internet2 and ESnet high-performance networks infrastructure in the US to deploy its Web-based system. By now, there are 70 reflectors that manage the traffic flow, as depicted in Figure 1.

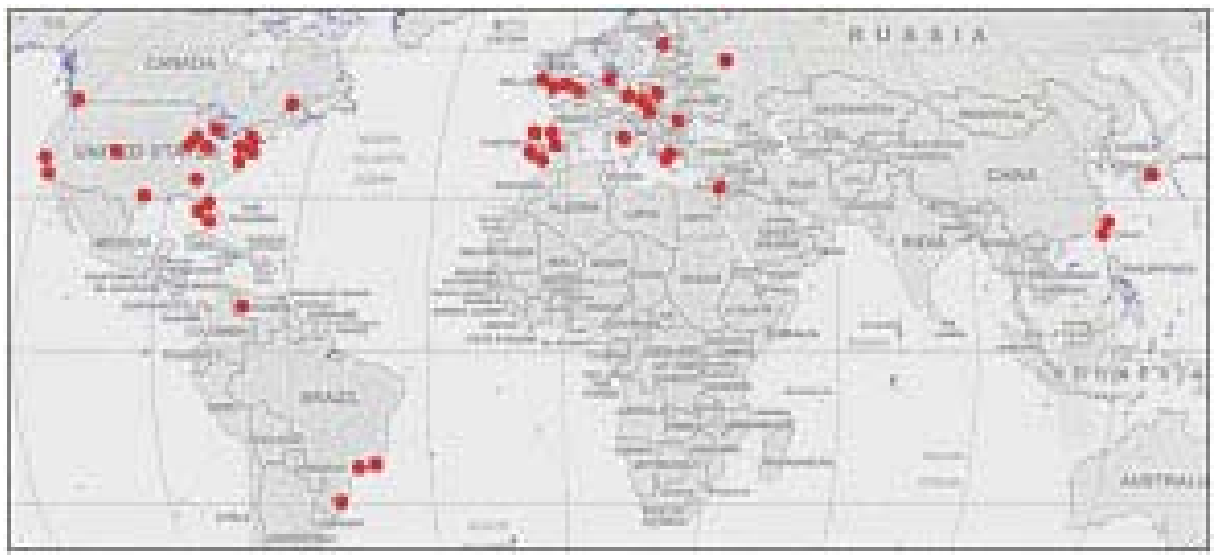

Figure 1 – VRVS Reflectors Topology

VRVS is also involved in the Global Grid Forum initiative, where new services are starting to be developed, based on the Open Grid Services Infrastructure (OGSI). Among these services, there are plans to monitor real-time activity of the system, oversee and manage the VRVS distributed network. This will make VRVS collaboration a persistent part of the overall Grid standards-based working environment that is currently being developed by the HENP community.

[1] MonALISA stands for Monitoring Agents in Large Integrated Services Architecture

## 2.1. VRVS Version 3

The VRVS site is designed to allow fast and intuitive navigation. Based on the "Virtual Rooms" concept, distributed users can meet in virtual spaces to collaborate. For a few years now, VRVS has provided a web based booking system where the participants can reserve meetings manually or through a "booking wizard", protect meetings with a password, check their latest bookings or the daily booked ones, etc. Also, during the meetings, a chat and a sharing service are available for easier collaboration.

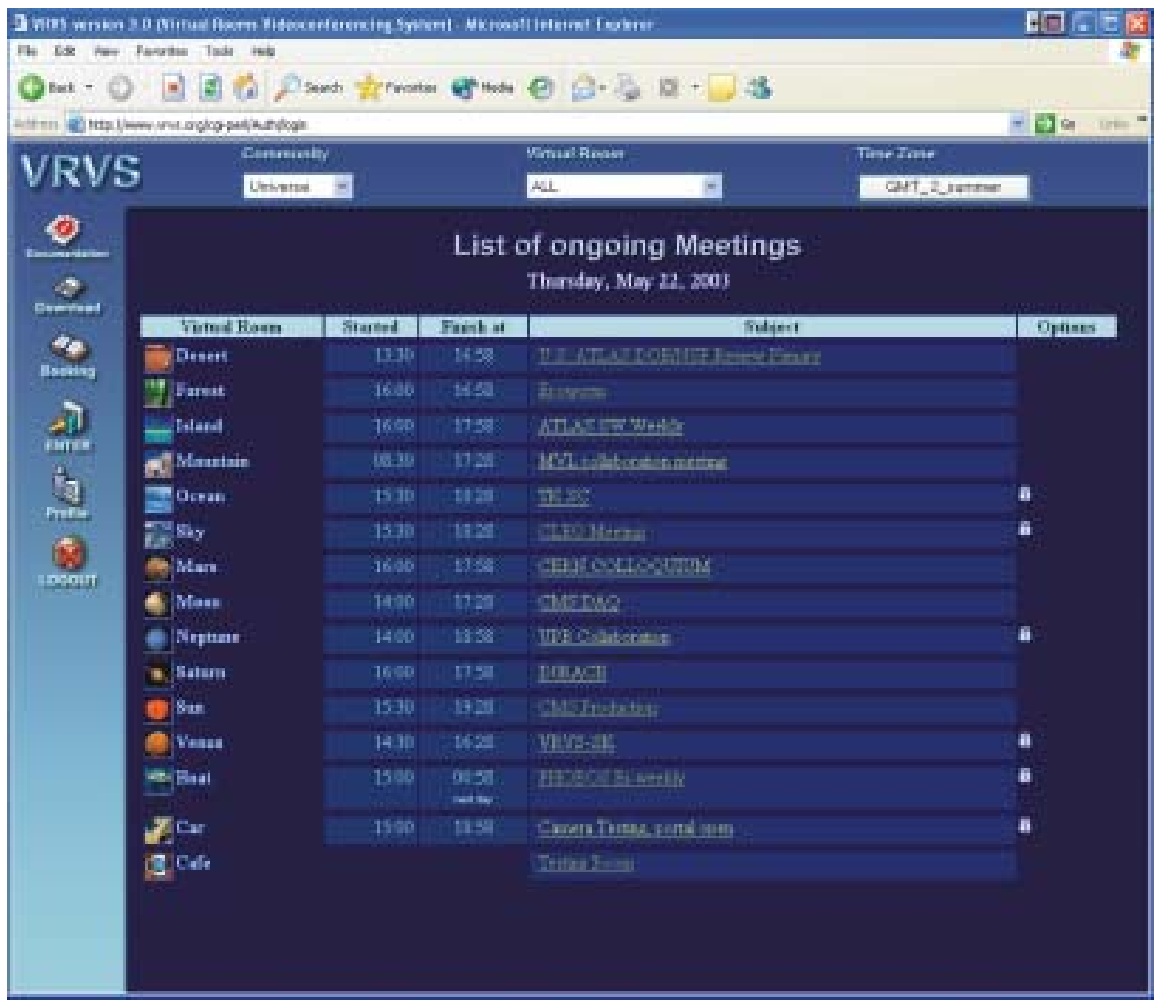

Figure 2 – Typical Ongoing VRVS Meetings

Since mid-February 2003, a new version of VRVS (version 3) has been successfully deployed. At the present time, more than 4,000 users from 81 countries have registered and used the new version of VRVS. There are currently 70 reflectors to interconnect and manage the traffic flow worldwide. Recently, new reflectors have been installed in Belgium, Brazil, France and China. Also, 5 new reflectors have been deployed in several US universities using the Internet2 Backbone.

In addition to improving dramatically the scalability and usability, the new release brings new features and capabilities. Tunneling between peer reflectors, NAT support, integration of the VNC sharing service through the reflectors, improvements in the booking system and user profiles, Mac OS X support are part of the new features available. In addition, the world time zones are now automatically managed by the system, so each user will see the time in their local time zone, regardless of their location

All these aspects have been integrated gradually and smoothly, driven by the user needs and demands. For a few months both VRVS 2.5 and 3 versions ran in parallel. During that time, the users were able to test the new system, get used to it and to its features. This transition let us fix small problems that were detected and reported. At the end of this transition period, the switch was completed and the result was a near-perfect migration of thousands of users from two completely different systems, without any major problem.

## 2.2. The VRVS AG gateway

With VRVS 3, the VAG (VRVS AG Gateway; or Virtual Access Grid) has been improved based on users experience and feedback. The new VAG has full connectivity to Access Grid and full functionality. VAG give users a minimal learning curve, and supports combined multicast-and-unicast collaborative sessions. The VAG has been shown to support a full Access Grid session on a laptop, consuming a few Mbps of bandwidth (or less, under user control), and can run over 802.11a or 802.11g wireless networks without packet loss. VAG reflectors have been installed in Internet2, and at Argonne National Laboratory, and will be deployed on institutional AG nodes as needed, based on users' requests. A VAG reflector is functionally identical to other VRVS reflectors and is very easy to configure.

To connect to Access Grid Virtual Venues or any multicast videoconferencing, VRVS users only need to login to VRVS 3, and within 5 intuitive clicks, the user is ready for a collaborative session that includes both AG and other VRVS participants. VRVS users have the maximum flexibility to choose from UCL Mbone, OpenMash Mbone, H.323, SIP, QuickTime, JMF (Java Media Framework) on various platform including Linux, Windows and Mac OS X. The audio transcoder has been improved to transcode AG linear L16-16-Mono to the ITU H.323 standard G.711 μ-Law.

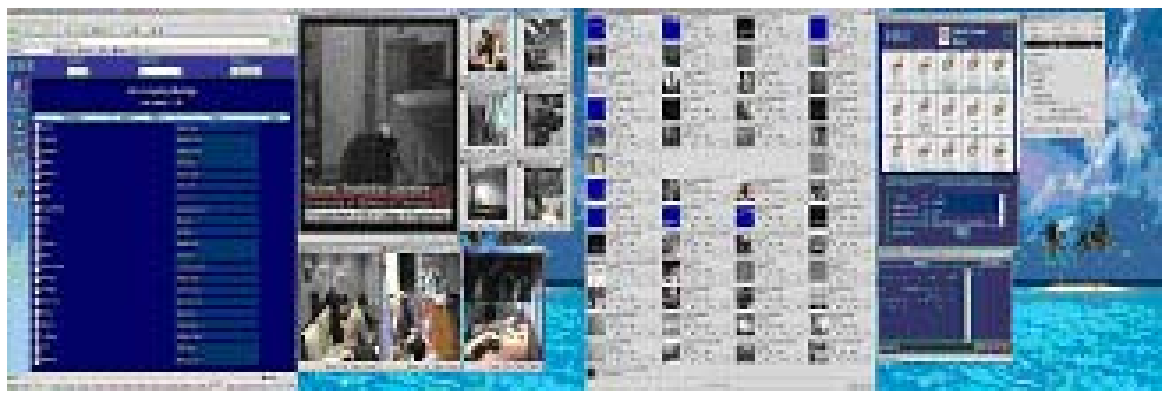

Figure 3 – VRVS-AG Gateway

An audio mixer feature is implemented to support H.323 audio mixing and avoid blocked video because of a noisy site injecting noise into the session. The new VAG also supports a useful range of video modes, particularly to accommodate VRVS users with limited local network and/or CPU power capacity. Specifically, VRVS provides four video modes: (1) Voice switched - the default mode for H.323 clients, receiving one video stream at a time; (2) Timer switched - one browses through all the video based on preset timer, receiving one video stream at a time; (3) Selected Streams – the default mode for Mbone clients. Click among the video participants to view selected video streams (one or several streams are available), which is useful for limited bandwidth network connections and/or legacy low-power local computing systems; and (4) All Streams - Mbone will receive all the video streams subscribed to the virtual venue multicast address. This is the best mode for full interactivity, if the network will support the data flow.

Development of a personal or small group VAG setup prototype is underway at Caltech. VRVS users can get 6400 X 1200 pixels of display space by driving 4 screens with a dual Xeon 2.8 Ghz PC server, for under $8,000. This will support a next-generation analysis environment that includes the usual analysis applications, Grid Views, multiple VRVS windows or a VAG session.

## 3. VRVS AS A UNIQUE SYSTEM

The next generation of collaborative environments will evolve to be more scalable and reliable, integrate new emerging standards, support new products, and provide very high-end quality, state-of-the-art efficiency in terms of quality per bandwidth.

At the present time, VRVS supports inter-connectivity among the most popular videoconferencing tools running on different operating systems either over unicast or multicast networks. The reflectors' backbone provides a pure software-based MCU with peer to peer structure using a sophisticated real time multipoint algorithm with low cost and maintenance. In this area, one of the future improvements we are already working on is the real time monitoring of reflectors, dynamic interconnection of reflectors based on detection of problems and system optimization, and alarm notification tools. All these features are based in the Caltech MonALISA project later described in this article. At a later date, we will also consider migrating some of these characteristics to the Open Grid Services Architecture.

All these aspects combined are transforming the VRVS system into the first of a new class of large scale distributed systems with real time constraints, unique in the production market.

## 4. NEXT DEVELOPMENTS

Several new developments are planned to the next versions of VRVS. In the meantime, some short term features/enhancement of the current production system will be considered as minor releases of version 3.x.

We are currently working on the integration of privileged users as sessions' chairmen, to mute/unmute the video/audio of any participant in real time, decide who the speaker is and which video should be received by remote participants.

In addition, we are developing a troubleshooting wizard that will run on the users' machines to determine if the system is correctly configured for VRVS use, and will help to diagnose problems.

Other enhancements are the full integration of the SIP protocol, automatic upgrade of VRVS reflectors using encryption and authentication, full compliance with IPv6, encryption of communications among reflectors, development of new high end videoconferencing services based on MPEG and/or HDTV, start the interoperability of handheld devices and provide powerful monitoring and tracking tools of VRVS usage (per user/ per reflector, per hour, etc.)

### 4.1. MonALISA Integration

Following the release of VRVS 3, the major new development has been the integration of the MonALISA monitoring service into the VRVS system. MonALISA was adapted and deployed on the VRVS reflectors. Dedicated modules to interact with the VRVS reflectors were developed: to collect information about the topology of the system; to monitor and track the traffic among the reflectors and report communication errors with the peers; and to track the number of clients and active virtual rooms. In addition, overall system information is monitored and reported in real time for each reflector: such as the load, CPU usage, and total traffic in and out.

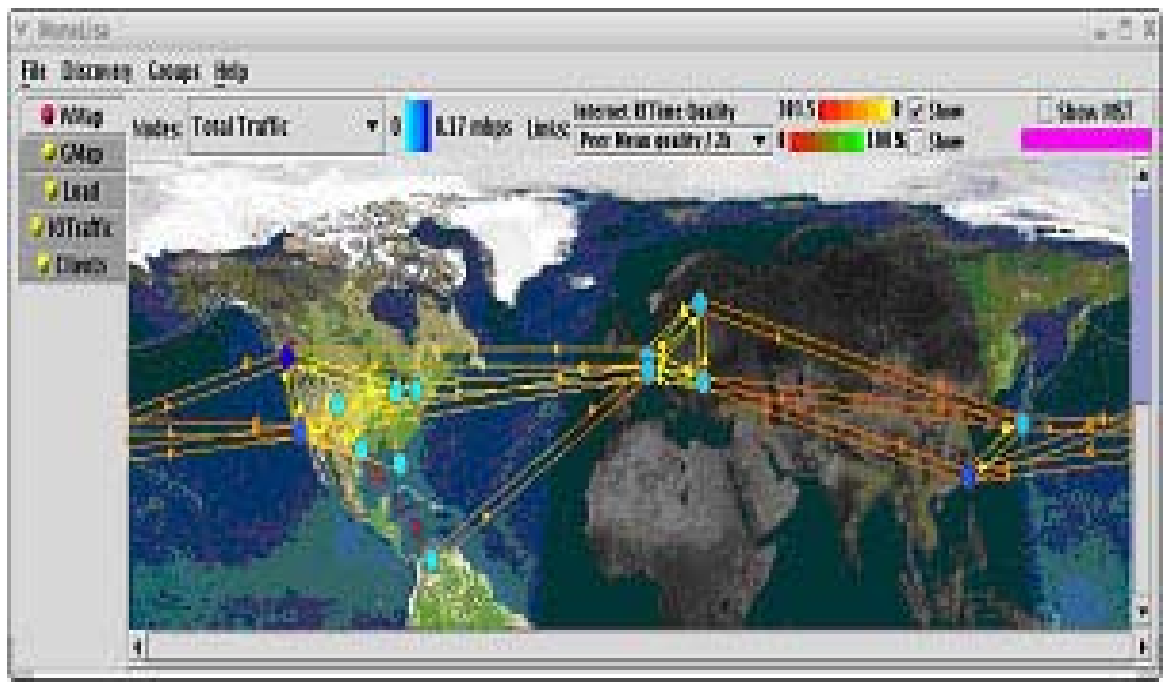

Figure 4 – VRVS-backbone in real time.

For each VRVS reflector, a MonALISA service is running as a registered JINI service. For the VRVS version the MonALISA service is used with an embedded Database, for storing the results locally, and runs in a mode that aims to minimize the reflector resources it uses (typically less than 16MB of memory and less than 1% of the system load.)

A dedicated GUI for the VRVS version was developed as a java web-start client.

This GUI provides real time information dynamically for all the reflectors which are monitored, as illustrated in Figures 4 and 5.

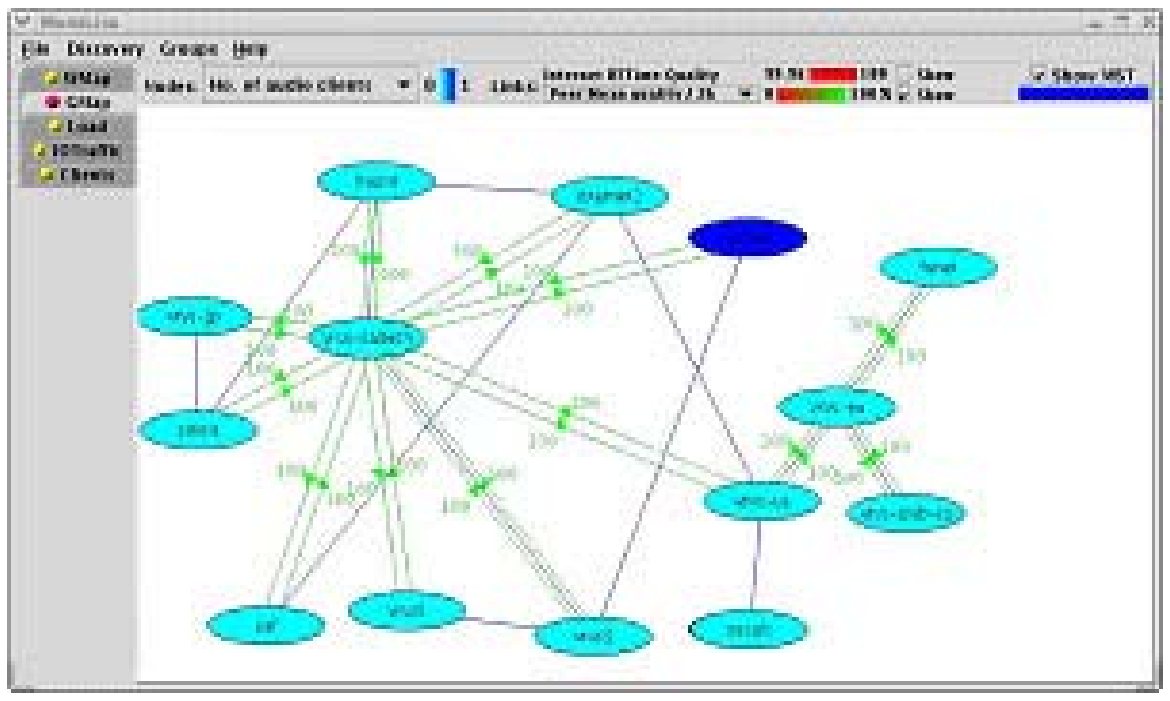

Figure 5 – Real time connectivity among peers.

If a new reflector is started it will automatically appear in the GUI and its connections to its peers will be shown. Filter agents to compute an exponentially mediated quality factor of each connection are dynamically deployed to every MonALISA service, and they report this information to all active clients who are subscribed to receive this information. A maximum flow algorithm has been implemented to optimize the way the reflectors are connected. In the graphical view the maximum flow path is shown with dark lines.

The subscription mechanism allows one to monitor in real time any measured parameter in the system. Examples of some of the services and information available, visualizing the number of clients and the active virtual rooms, the traffic in and out of all the reflectors, as well as problems such as lost packets between reflectors are shown in Figure 6.

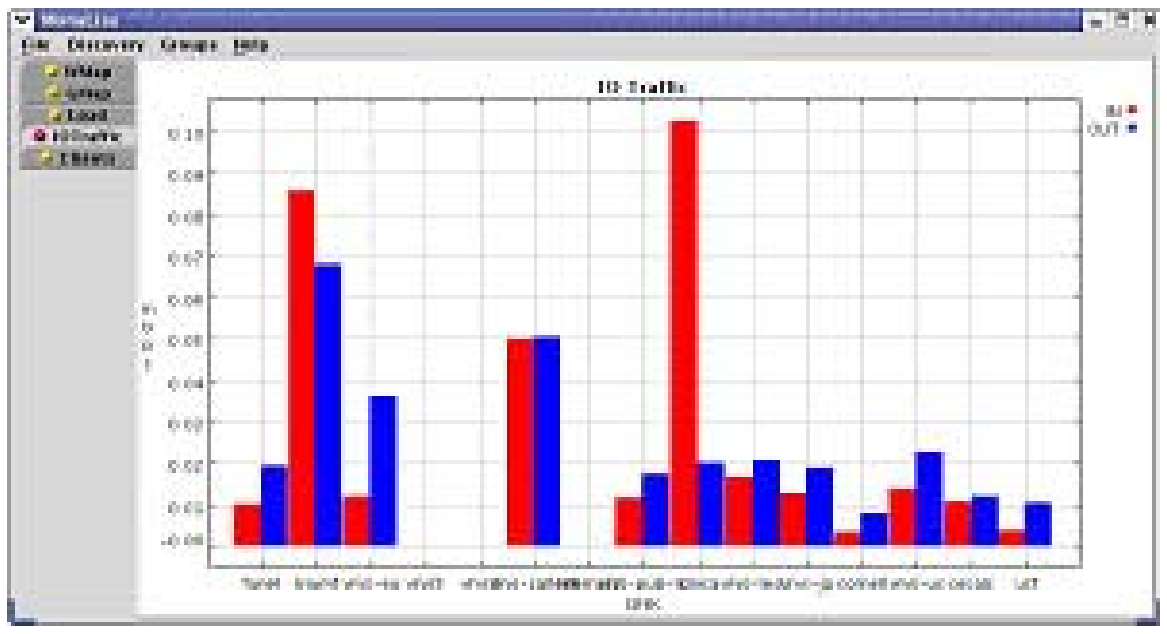

Figure 6 – Real time reflectors' info.

In addition to dedicated monitoring modules and filters for the VRVS system, agents have been developed to supervise the running of the VRVS reflectors automatically. This will be particularly important when scaling up the VRVS system further. In case a VRVS reflector stops or does not answer the monitoring requests correctly, the agent tries to automatically restart it. If this operation fails twice the Agent will send an email to a list of administrators. These agents are the first generation of modules capable of reacting and taking well defined actions when errors occur in the system.

We are developing agents able to provide an optimized dynamic routing of the videoconferencing data streams. These agents require information about the quality of alternative connections in the system and they solve a minimum spanning tree problem to optimize the data flow at the global level. These agents are capable to take system actions and may be dynamically loaded and digitally signed by developers with trusted certificates (for security reasons).

## 4.2. Pocket VRVS

Another new development that VRVS Team has currently underway is the pocket PC platform support.

As with all computer technologies today, handheld computers are becoming more and more powerful. They are now able to run the software codecs needed to process the real-time audio and video streams needed for a fluid videoconference. Wireless connections are also becoming faster. With 802.11a (and the upcoming 802.11g) able to send 54 Mbps (up to 24 Mbps in practice) as opposed to 802.11b which can send only 11 Mbs (to 4-5 Mbps in practice) sending full audio and video streams over a wireless connection is now practicable. More cameras for handheld computers are coming on the market every day. These cameras work at faster frame rates and higher resolutions than before, and more videoconferencing clients for handheld computers are becoming available. There are now H.323 and SIP clients for handheld computers which could be integrated into VRVS. This would make VRVS more of a ubiquitous tool for physics collaborations at work.

A prototype of a VRVS audio/video client that runs on a Pocket PC has been implemented (Figure 7.) It supports the standard H.261 CIF video and G.711 μ-Law audio. We still need to construct interfaces to run on the smaller screen space but include all the controls we have for all the clients (meeting scheduler, chat, etc).

Especially vital for handheld videoconferencing are the video controls. It is only practical to display one video at a time because of the limited screen space of a handheld computer (usually just 'CIF' size, namely 288 by 352 pixels). Being able to choose easily which video of the participants in the meeting to view (using shortcuts) could be important in the future, especially when using a handheld Virtual Access Grid in which each venue can hold 30 or more videos.

To help enable seamless collaboration we can now allow mobile individuals to connect as a group in a meeting. Using wireless handheld computers with audio and video capabilities anyone in a group will be able to participate in a meeting at any time anywhere.

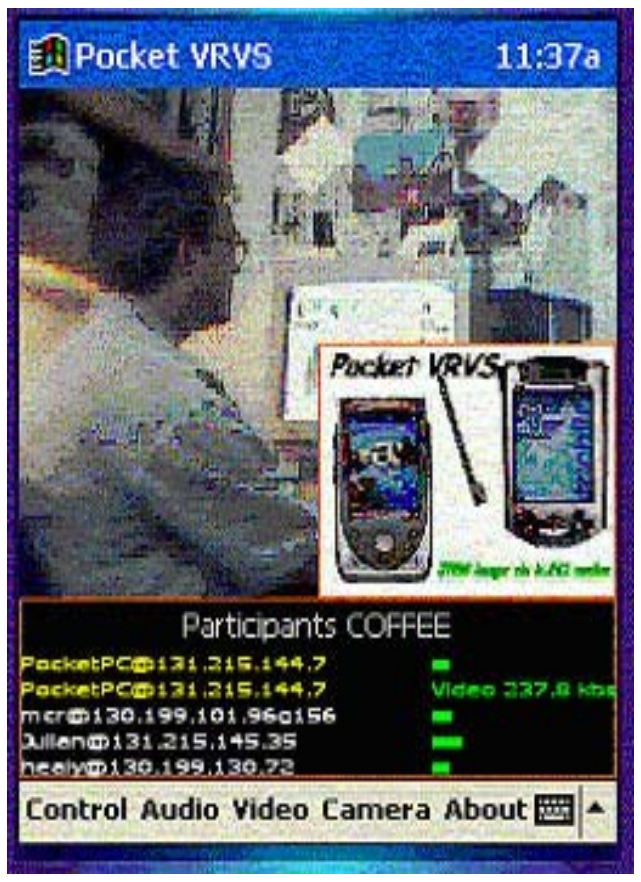

Figure 7 – VRVS in a Pocket PC.

The VRVS Team will continue to develop and expand the system with new available technologies, adding, among others features, the ones described in this document to provide the most powerful tool available for collaborative environments. This tool provides a unique, independent, flexible and scalable platform, for a professional collaborative experience.